\documentclass[fleqn,aps,prl]{revtex4}

\usepackage{amsmath,amssymb}
\usepackage{graphicx,color}
\usepackage{bm}
\usepackage{url} %
\usepackage{physics}
\usepackage{mathtools}
\usepackage{color}

\usepackage[normalem]{ulem}
\newcommand{\stkout}[1]{{\ifmmode\text{\sout{\ensuremath{#1}}}\else\sout{#1}\fi}}

\newcommand{\figref}[1]{\figurename~\ref{#1}}

\newcommand{\ii}{\mathrm{i}}
\newcommand{\re}[1]{\operatorname{Re}\left( #1 \right)}

\renewcommand{\ip}[2]{\left\langle #1, #2 \right\rangle}

\newcommand{\conj}[1]{\overline{#1}}
\newcommand{\adj}[1]{{#1}^*}

\newcommand{\lt}[1]{\left\langle #1\right \rangle_t}

\newcommand{\kdeath}{\kappa_\mathrm{q}}
\newcommand{\kc}{\kappa_\mathrm{c}}
\newcommand{\kci}{\kappa_\mathrm{c}}

\newcommand{\thdeath}{\theta_\mathrm{q}}
\newcommand{\cth}{c_\mathrm{q}}

\begin{document}
\title{Two distinct transitions in a population of coupled oscillators with turnover: desynchronization and stochastic oscillation quenching}
\author{Ayumi Ozawa}
\email[corresponding author: ]{aozawa@g.ecc.u-tokyo.ac.jp}
\author{Hiroshi Kori}
\affiliation{Graduate School of Frontier Sciences, The University of Tokyo, Chiba 277-8561, Japan}
\date{\today}

\begin{abstract}
Synchronization, which is caused by mutual coupling, and turnover, which is the replacement of old components with new ones, are observed in various open systems consisting of many components. Although these phenomena can co-occur, the interplay of coupling and turnover has been overlooked. Here, we analyze coupled phase oscillators with turnover and reveal that two distinct transitions occur, depending on both coupling and turnover: desynchronization and what we name stochastic oscillation quenching. Importantly, the latter requires both the turnover and coupling to be sufficiently intense.
\end{abstract}
  \maketitle

\paragraph{Introduction}
 The emergence of order in open systems comprising many interacting units is widely observed in nature. In particular, mutual synchronization is observed in a wide range of coupled oscillator systems \cite{Pikovsky2001book}, from biological systems \cite{Glass2001,winfree01} to social \cite{Neda2000} and artificial ones \cite{Soriano2013,Wiesenfeld1996}. Another phenomenon broadly observed in open many-body systems is turnover owing to the addition and removal of components. Examples include the protein and cell turnover in biological systems \cite{Rolfs2021,Pellettieri2007}. Other examples are found in social \cite{Burnett2011,Gualdi2015} and bio-inspired chemical systems \cite{Sugiura2016}. Further, a growing population may be effectively modeled as a system with turnover when the growth causes the dilution of components \cite{Zwicker2010, YuWood2015}.
 These two phenomena, mutual synchronization and turnover, may manifest in the same system, and their time scales are not always clearly separated. For example, KaiC proteins in a cyanobacterial cell exhibit a collective rhythm of phosphorylation and dephosphorylation with a period of approximately $24$h, and their average half-life has been estimated to be approximately $10$h \cite{Imai2004}.

Recent studies have revealed that turnover can 
deteriorate the collective oscillation; it has been shown numerically \cite{Zwicker2010} and experimentally \cite{Teng2013} that the synchronous phosphorylation-dephosphorylation cycle of KaiC proteins loses robustness when the turnover rate is a sufficiently large constant. Moreover, Ref.~\cite{Zwicker2010} has suggested that their collective oscillation disappears when the turnover rate is further increased. However, little attention has been paid to the synergistic effect of the interaction among oscillators, which is essential for mutual synchronization, and the turnover. In particular, these previous studies lack analyses on how the effect of the turnover changes as the properties of the coupling are varied.

In this regard, we study a simple model of coupled phase oscillators with turnover and show that 
their collective oscillation disappears 
via two distinct transitions depending on both the coupling and turnover. For sufficiently small coupling strengths, the collective oscillation is lost via desynchronization \cite{Kuramoto1984} as the turnover rate increases, while for stronger coupling strengths, we may observe what we refer to in this study as \emph{stochastic oscillation quenching} (SOQ), which can be interpreted as a stochastic analog of oscillation quenching \cite{Zou2021}. Interestingly, SOQ may be induced not only by increasing the turnover rate but also by strengthening the interaction among oscillators. 
Thus, this extinction of the collective oscillation is 
caused by a synergistic effect of the turnover and coupling. Our model is based on the Kuramoto model \cite{Kuramoto1975,Kuramoto2019,Strogatz2000,Acebron2005}, which has been successfully applied to investigate synchronization in various systems \cite{Kuramoto1984, winfree01, Pikovsky2001book, Wiesenfeld1996,Zhai2004,Kiss2002,Neda2000}, and incorporates the effect of the turnover as stochastic resetting \cite{Evans2020, Nagar2023}. The tractability of the model enables us to obtain transition curves.

\paragraph{The Kuramoto model with turnover}
The dynamics of $N$ identical Kuramoto oscillators are given as follows \cite{Kuramoto1984,Kuramoto1975}:
\begin{align} %
 \dv{\theta_i}{t} = \omega +
 \frac{\kappa}{N}\sum_{j=1}^{N}\sin\left(\theta_j - \theta_i\right),\label{eq:kuramoto}
\end{align}
where $\theta_i$ $(i=1,2,\ldots,N)$ is the phase of the $i$th oscillator, $\omega \neq 0$ is the natural frequency of the oscillators, and $\kappa \geq 0$ represents the coupling strength. 
  The Kuramoto model exhibits mutual synchronization and has the advantage of ease of analytical treatment \cite{Acebron2005}. However, modeling the turnover of such oscillators in a tractable manner is a non-trivial problem because the removal and addition of oscillators involve changes in the number of variables comprising the dynamical system. We circumvent this difficulty by constructing a model as follows.

Suppose one of the oscillators is randomly chosen and replaced by a new oscillator. This event is equivalent to resetting the phase of the selected oscillator to that of the newly added oscillator.
Hence, as a model for oscillators with turnover, we adopt a system that involves phase resetting. 
Specifically, we consider a population of Kuramoto oscillators where each oscillator experiences a reset event with probability $\alpha \dd t$ during an infinitely short time width $\dd t$. In other words, the resetting events are such that $\alpha N$ oscillators are expected to be substituted per unit time.
Such dynamics can be described by the following It\^{o} stochastic differential equation with jumps:
\begin{align}
  \dd \theta_i&= \left[\omega 
        + \frac{\kappa}{N}\sum_{j=1}^{N}\sin\left(\theta_j - \theta_i\right)
      \right] \dd t  \notag \\
      &+\left( -\theta_i + \phi_i \right) \dd P_i(\phi_i;\alpha),
       \label{eq:sde}    
\end{align}
  where $d P_i(\phi_i; \alpha)$ is the differential of a marked Poisson process $P_i(\phi_i; \alpha)$ with intensity $\alpha$, describing the resets of the $i$th oscillator. The number of resetting events during a period $\Delta t$ obeys the Poisson distribution with intensity $\alpha \Delta t$, and the value of the random variable $\phi_i$, which is called the mark \cite{Baddeley2007} of the Poisson process $P_i$, is drawn from a distribution $f(\phi)$ upon each resetting event. 
  When a resetting event of the $i$th oscillator occurs, $\theta_i$ changes by $-\theta_i + \phi_i$. 
  Thus, $\phi_i$ is a random variable corresponding to the phase of the $i$th oscillator just after its reset. 
  Hereinafter, we refer to $\alpha$ as the turnover rate. 
  Note that the resetting events occur independently, i.e., 
  the counting processes $\left\{ P_i(\phi_i;\alpha) \right\}_{i=1}^{N}$ are independent  
  and the values of the marks at different resetting events are independently drawn from $f(\phi)$.

\begin{figure}
  \includegraphics{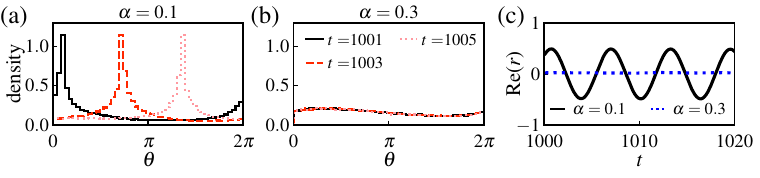}
    \caption{
      An increase in the turnover rate $\alpha$ quenches the collective oscillation. 
      (a), (b) Snapshots of the phase distribution for $t=1001$, $t=1003$, and $t=1005$, obtained from the numerical simulation of Eq.~\eqref{eq:sde} by the Euler–-Maghsoodi method \cite{Hanson2007}. These panels share the same legend.
      (a) The propagation of the distinct peak of the distribution is observed for $\alpha=0.1$. (b) For $\alpha=0.3$, the distribution is steady and flatter compared to that for $\alpha=0.1$. 
      (c) Time series of the real part of the complex order parameter $r$. The black solid and blue dotted curves illustrate the dynamics of $\re{r}$ for $\alpha=0.1$ and $\alpha=0.3$, respectively. The oscillation observed for $\alpha=0.1$ disappears for $\alpha=0.3$.  
      In all the simulations, the initial state is the uniform distribution. Other parameters are $\omega=1$, $\kappa=0.3$, $\sigma=0.5$, and $N=50000$.
    }
    \label{fig:dynamics}
\end{figure}

    Equation \eqref{eq:sde} was simulated \footnote{See Supplemental Material at [URL will be inserted by publisher] for details.} using Euler-Maghsoodi method \cite{Hanson2007}.
Consistent with previous work \cite{Zwicker2010}, the increase in the turnover rate $\alpha$ in Eq.~\eqref{eq:sde} causes the extinction of the macroscopic oscillation. The snapshots of the phase distribution in \figref{fig:dynamics}(a,b) indicate that the distribution for $\alpha=0.1$ has a sharp peak and changes with time, whereas that for $\alpha=0.3$ is almost uniform and steady. In the numerical simulation, we adopted $f(\phi)$ with the form of a Poisson kernel:
\begin{align}
    f(\phi) = \frac{1}{2 \pi}\frac{1-\sigma^2}{1 - 2 \sigma \cos \phi + \sigma^2},
    \label{eq:poisson_kernel}
\end{align}
where $\sigma$ determines the sharpness of $f(\phi)$; $f(\theta) = 1/(2\pi)$ for $\sigma=0$ and $f(\theta) \to \delta(\theta)$ as $\sigma \to 1$. 
Similar results were obtained for other unimodal distributions (not shown here).

Let us quantify the macroscopic oscillation by introducing the complex order parameter $r$, Kuramoto order parameter $R$, and mean phase $\Theta$ as follows \cite{Kuramoto1984, Strogatz2000}:
\begin{align}
    r(t) = R(t)e^{\ii\Theta(t)} \coloneqq \frac{1}{N}\sum_{j=1}^{N} e^{\ii \theta_j(t)}. \label{eq:op}
\end{align}
According to this definition, $R$ reflects the coherence of the phases; $R=0$ when the phases are uniformly distributed, and $R=1$ when all oscillators have the same phase. Note that the system without turnover, Eq.~\eqref{eq:kuramoto}, has a stable synchronized oscillatory solution with $R(t)=1$ and $\Theta(t)=\Theta(0)+\omega t$. The time series of $\re{r}$ oscillates for $\alpha=0.1$, whereas it remains constant for $\alpha=0.3$; see Fig.~\ref{fig:dynamics}(c).

\begin{figure}
  \includegraphics{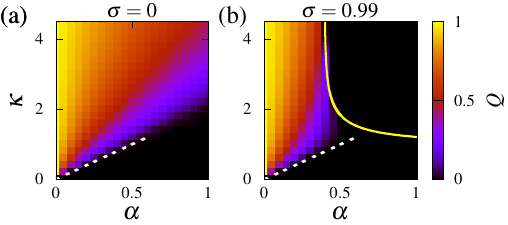}
  \caption{
    Two transition curves explain the disappearance of the collective oscillation. The index $Q$ of the intensity of the collective oscillation, defined by Eq.~\eqref{eq:Q}, is averaged over $5$ realizations of the Poisson processes and shown in color scale for (a) $\sigma=0$ and (b) $\sigma=0.99$. When $\alpha$ and $\kappa$ are sufficiently small, the critical coupling strengths below which $Q$ vanishes coincide with the desynchronization transition curve, $\kappa=\kci$, which is shown via a white dotted line in both panels. 
    In contrast, when $\kappa$ is sufficiently large, the boundary of the region of collective oscillation depends on $\sigma$; in (a), the boundary depends on $\kappa$ almost linearly even for $\alpha \sim 1$, while in (b), another transition curve in addition to $\kappa=\kc$ is needed to fully describe the boundary. The supplementary curve obtained by analyzing SOQ is illustrated by a yellow solid curve. The initial condition is set to the uniform distribution. Other parameters are $\omega=1$ and $N=2000$.
    }
  \label{fig:Q}
\end{figure}

We define the intensity of the macroscopic oscillation as the fluctuation of $r$:
\begin{align}
   Q \coloneqq \sqrt{ \lt{\left| r - \lt{r} \right|^2 }}, \label{eq:Q}
\end{align}
where $\lt{}$ denotes the long-time average. Note that $Q=0$ if $r$ is constant. As shown in Fig.~\ref{fig:Q}, the dependence of $Q$ on phase resetting changes qualitatively as the coupling strength $\kappa$ varies. For sufficiently small $\kappa$, the value of $Q$ is hardly affected by the width $\sigma$ of the distribution $f$ and gradually decreases as $\alpha$ increases. In contrast, for larger values of $\kappa$, $\sigma$ also affects $Q$; when $\sigma \simeq 0$, $Q$ gradually decreases as $\alpha$ increases (Fig.~\ref{fig:Q}(a)), whereas $Q$ suddenly drops to near $0$ when $\sigma \simeq 1$ (Fig.~\ref{fig:Q}(b)).
  The choice of initial conditions does not seem to affect the results; completely synchronized initial conditions, $\theta_1(0) = \theta_1(0)  \dots =\theta_N(0) = 0$, yielded almost the same results as in \figref{fig:Q}, where the initial phase distribution was set to the uniform distribution. Additionally, no hysteresis was observed when the value of $\alpha$ was varied backward and forward near the transition point (not shown here).

\paragraph{Mean-field approximation}
\label{sec:PDE}
To analyze these transitions, 
let us reduce the system by employing a mean-field approximation in the same manner 
as in \cite{Crawford1999}. 
First, let $p(\vb*{\theta}, t)$ and $p_i(\theta_i, t)$ be the probability density function for $\vb*{\theta}\coloneqq(\theta_1, \theta_2,\ldots, \theta_N)$ at time $t$ and its marginalization over all phase variables except $\theta_i$, respectively. 
From Eq.~\eqref{eq:sde}, the time evolution of $p_i(\theta_i, t)$ is given by
\begin{align} %
 \pdv{p_i}{t} = 
 - \pdv{}{\theta_i}\left[\int_{S_i} p(\vb*{\theta}, t)v_i(\vb*{\theta})\dd \vb*{\theta}_i\right] + \alpha f- \alpha p_i, \label{eq:forward_k}
\end{align}
where
$\vb*{\theta}_i \coloneqq \left( \theta_1, \theta_2,\ldots, \theta_{i-1}, \theta_{i+1},\ldots, \theta_N \right)$, 
$\int_{S_i} \cdot \dd \vb*{\theta}_i \coloneqq  \int_0^{2\pi}\cdots \int_{0}^{2\pi} \cdot \dd \theta_1 \dd \theta_2 \cdots \dd\theta_{i-1}\dd \theta_{i+1}\cdots \dd \theta_{N}$, and 
$v_i(\vb*{\theta}) \coloneqq \omega + \kappa N^{-1}\sum_{j=1}^{N} \sin\left( \theta_j - \theta_i \right)$ \cite{Hanson2007}. 
  In Eq.~\eqref{eq:forward_k}, the effect of stochastic resetting is described by the second and third terms on the right-hand side of the equation. Because the stochasticity of the system arises only from jump processes, the equation has a linear term $-\alpha p_i$ rather than terms with $\pdv*[2]{p_i}{\theta_i}$.
 
Now we invoke the mean-field approximation; we assume $p_i(\theta_i,t)p_j(\theta_j,t) - p_{i,j}(\theta_i, \theta_j, t)\to 0$ and $p_{i}(\theta_i, t) - p_{j}(\theta_j, t) \to 0$ as $N \to \infty$ for any $i \neq j$, where $p_{i,j}(\theta_i,\theta_j, t)$ is the joint probability distribution function for $\theta_i$ and $\theta_j$.
Then, from Eq.~\eqref{eq:forward_k}, the time evolution equation of the phase distribution $p(\theta, t)$ is obtained as
\begin{align}
  &\pdv{p}{t}= 
  - \pdv{}{\theta}\left[pv\right]+ \alpha f- \alpha p,
  \label{eq:forward_k_meanfield} \\
  &v=v(\theta, t) \coloneqq \omega 
  + \kappa \int_0^{2\pi} p(\tilde{\theta},t)\sin(\tilde{\theta} - \theta) \dd \tilde{\theta}.
\end{align}
In the following, we set $\omega=1$ without loss of generality by normalizing $t$, $\kappa$, and $\alpha$ by $\omega$. 

\paragraph{Desynchronization}
When $\alpha=0$, Eq.~\eqref{eq:forward_k_meanfield} has the uniform steady solution $p^{(0)}(\theta)\coloneqq 1/(2\pi)$. This solution corresponds to the desynchronized state in the sense that the cohesiveness of the phases is completely lost. 
For $\alpha \neq 0$, $p^{(0)}(\theta)$ is no longer a solution of Eq.~\eqref{eq:forward_k_meanfield}; however, \figref{fig:dynamics}(b) suggests the existence of a steady solution obtained by slightly deforming $p^{(0)}(\theta)$. 
The stabilization of such a solution, formally analyzed below, explains the extinction of the collective oscillation observed for small $\kappa$ and $\alpha$ in Fig.~\ref{fig:Q}.

Let $\rho$ be the deviation from a steady solution
$\hat{p}$, i.e., $\rho = p - \hat{p}$.
Linearizing Eq.~\eqref{eq:forward_k_meanfield} around $\hat{p}$ yields
\begin{align}%
\pdv{\rho}{t} &\simeq -\pdv{}{\theta}\left\{ \hat{p} \mathcal{U}\left[\rho\right]+
\left( 1+\mathcal{U}\left[\hat{p}\right] \right)\rho\right\} - \alpha \rho \eqqcolon \mathcal{L}[\rho], \label{eq:def_L} 
\end{align}
where the linear functional $\mathcal{U}$ is defined by
\begin{align}
  \mathcal{U}[\rho](\theta) &\coloneqq \kappa \int_0^{2\pi}\sin\left( \tilde{\theta} - \theta \right)\rho(\tilde{\theta}) \dd \tilde{\theta}.    
\end{align}
Assume further that $\hat{p}$ and $\mathcal{L}$ are expanded as follows for sufficiently small values of $\alpha$:
\begin{align}
    \hat{p}(\theta;\alpha,\kappa) &= \hat{p}^{(0)}(\theta) + \sum_{l=1}^{\infty} \alpha^l p^{(l)}(\theta;\kappa), \label{eq:p_expansion}\\
    \mathcal{L} &= \sum_{l=0}^{\infty} \alpha^l \mathcal{L}^{(l)}. \label{eq:L_expansion}
\end{align}
The expression for $\mathcal{L}^{(l)}$ is obtained by 
inserting Eqs.~\eqref{eq:p_expansion} and \eqref{eq:L_expansion} into Eq.~\eqref{eq:def_L} and collecting the terms of $\order{\alpha^l}$.
It is straightforward to find that $\mathcal{L}^{(0)} u_m^{(0)}(\theta) = \lambda_m^{(0)} u_m^{(0)}(\theta)$ holds for any integer $m$, where $u_m^{(0)}(\theta) = e^{\ii m \theta}$ and $\lambda_m^{(0)} = -\ii m + \frac{\kappa}{2}\delta_{|m|,1}$. 

Now, we define an inner product $\ip{\cdot}{\cdot}$ of smooth $2\pi$-periodic functions $z_1(\theta)$ and $z_2(\theta)$ as 
\begin{align}
    \ip{z_1}{z_2} \coloneqq \frac{1}{2\pi}\int_0^{2\pi} z_1(\theta)\conj{z_2(\theta)} \dd \theta,
\end{align}
where $\conj{z_2(\theta)}$ is the complex conjugate of $z_2(\theta)$. The adjoint operator $\adj{\mathcal{L}^{(0)}}$ of $\mathcal{L}^{(0)}$ defined on this inner-product space satisfies $\adj{\mathcal{L}^{(0)}} u_m^{(0)} = \lambda_{-m}^{(0)} u_m^{(0)}$. Thus, $u_m^{(0)}(\theta)$ is an eigenvector of both $\mathcal{L}^{(0)}$ and $\adj{\mathcal{L}^{(0)}}$. 
This allows us to apply the Rayleigh--Schr\"{o}dinger perturbation theory \cite{Sakurai2011} to evaluate the eigenvalues of $\mathcal{L}$. Specifically,
we assume that the eigenvalues and eigenvectors of $\mathcal{L}$ can be expanded around those of $\mathcal{L}^{(0)}$:
\begin{subequations}
  \begin{align}
    \mathcal{L} u_m &= \lambda_m u_m,\label{eq:eigen_ansatz}\\
    u_m &= \sum_{l=0}^{\infty} \alpha^l u_{m}^{(l)},
    &\lambda_m = \sum_{l=0}^{\infty} \alpha^l \lambda_{m}^{(l)}.\label{eq:eigen_expansion} 
  \end{align}      
\end{subequations}
To ensure the uniqueness of the expansion, we also impose the orthogonality condition $\ip{u_m^{(0)}}{u_m^{(l)}} = \delta_{0,l}$.

Inserting Eqs.~\eqref{eq:L_expansion} and \eqref{eq:eigen_expansion} into \eqref{eq:eigen_ansatz} and extracting the terms of $\order{\alpha}$, we obtain
\begin{align}
    \lambda_m^{(1)} u_m^{(0)} = \mathcal{L}^{(0)} u_m^{(1)}+ \mathcal{L}^{(1)} u_m^{(0)} -
  \lambda_m^{(0)} u_m^{(1)}.\label{eq:expansion_order_1}
 \end{align}
Then, computing the inner product $\ip{\lambda_m^{(1)} u_m^{(0)}}{u_m^{(0)}}$ yields
  $\lambda_m^{(1)} =-1$. 
Hence, $\lambda_m = -\ii m + \delta_{|m|,1}\frac{\kappa}{2} - \alpha$ to the order of $\alpha$, and the maximum of the real parts of the eigenvalues exceeds $0$ when
\begin{align}
    \kappa \simeq 2 \alpha \eqqcolon \kci. \label{eq:kc}
\end{align}
Figure \ref{fig:Q} implies that, for small $\alpha$, the boundary on which $Q$ vanishes agrees with the curve $\kappa =\kci$, which is delineated by the white dotted curves. 
Consistent with the numerical simulation, $\kci$ depends on $\alpha$ but not on $\sigma$, the width of $f(\theta)$.

It should be noted that the stability analysis presented here is formal but not mathematically rigorous 
because (1) the validity of the perturbative calculation is assumed without any proof and (2) the continuous and residual spectra are ignored.
Nevertheless, the agreement with numerical simulation indicates the validity of the analysis.

\paragraph{Stochastic oscillation quenching}
The dependence of $Q$ on the turnover rate $\alpha$ for $\sigma=0$, shown in Fig.~\ref{fig:Q}(a), does not change qualitatively when $\kappa$ is increased. However, when $\sigma$ approaches $1$, the transition observed for large $\kappa$ is no longer explained by desynchronization discussed above; see Fig.~\ref{fig:Q}(b). 

\begin{figure}
\centering
\includegraphics{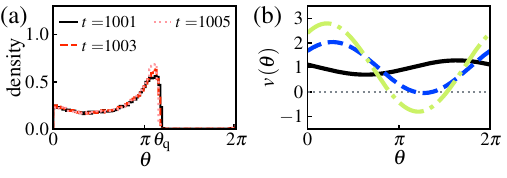}

 \caption{SOQ occurs for sufficiently large $\kappa$. (a) Snapshots of the histogram of the phases in the system near the transition point of SOQ. The line colors are the same as those in \figref{fig:dynamics}(a,b). The parameters are $(\kappa,\alpha,\sigma)=(2.5, 0.5, 0.99)$. (b) The SOQ transition involves the emergence of the zeros of velocity field $v(\theta)$. The blue dotted and green dot-dashed curves show $v(\theta)$ for $\kappa=2.2$, around which the SOQ transition occurs, and for $\kappa=3.0$, where the system is far beyond the transition point, respectively. The black solid curve is for $\kappa=0.3$, corresponding to the desynchronized state.}
 \label{fig:death}
\end{figure}

Numerical simulations indicate that the phase distribution after this transition is steady but far from uniform, as shown in Fig.~\ref{fig:death}(a). Furthermore, the velocity field
\begin{align}
  v(\theta)\coloneqq \omega + \kappa\int \sin(\theta'-\theta)p(\theta')\dd \theta'
   \label{eq:def_v}
\end{align}
 is qualitatively different from that of the desynchronized state; the transition involves the emergence of 
the zeros of $v(\theta)$, and these continue to exist when $\kappa$ or $\alpha$ is further increased. See the blue dashed and green dot-dashed curves in Fig.~\ref{fig:death} for $v(\theta)$ just after and far beyond the transition, respectively. In contrast, $v(\theta)$ in the desynchronized state, which is indicated by the black solid curve in Fig.~\ref{fig:death}, has no zero.

In deterministic systems, the zeros of the velocity field of the oscillators imply oscillation quenching, the phenomenon where individual oscillators cease their oscillation \cite{Zou2021}. Although our system involves stochastic resetting, we can make an analogy as follows. 
Consider a ``test oscillator,'' i.e., an oscillator that changes its phase according to $v(\theta)$ but does not affect the system nor undergo phase resetting.
If we incorporate it into the system, its phase changes toward $\thdeath$, a point at which $v(\theta)=0$ and $dv(\theta)/d\theta \leq 0$ hold, and then becomes almost static after initial transients. Thus, the steady distribution with which $v(\theta)$ has zeros implies the quenching of the test oscillator, and we refer to the realization of such a distribution as stochastic oscillation quenching (SOQ).

In the analyses of oscillation quenching, the condition that the time derivatives of the state variables vanish is often combined with the definition of order parameters to develop self-consistency arguments \cite{Fukai1994, Bi2014}. However, in the case of SOQ, such a condition can not be used because the phases keep changing toward $\thdeath$. Therefore, we use a condition regarding $v(\theta)$ instead. Specifically, in the following, a transition curve that explains numerical results for $\sigma \simeq 1$ is obtained by imposing the condition that $v(\theta)$ has exactly one zero in the limit $f(\theta) \to \delta(\theta)$.

Setting $\pdv{p}{t}=0$ in Eq.~\eqref{eq:forward_k_meanfield} in this limit and imposing $v(\thdeath)=\frac{\dd v}{\dd \theta}(\thdeath)=0$ yields
\begin{align}
  \pdv{}{\theta}\left\{ 
    p(\theta)\left[ 1-\cos(\theta_q - \theta) \right]\right\} + \alpha p(\theta) - \alpha \delta(\theta) = 0.
  \label{eq:ode_steady_death}
\end{align}
We seek a solution $p(\theta)$ such that $p(\theta)>0$ in $[0,\thdeath)$ and $p(\theta)=0$ in $[\thdeath, 2\pi)$ because 
the phase of a new oscillator continues to approach from $0$ to $\thdeath$  when SOQ occurs. This ansatz is consistent with the phase distribution obtained by numerical simulation for $\sigma=0.99 \simeq 1$, where most of the oscillators are located in the range $[0,\thdeath)$, as shown in \figref{fig:death}(a).
Noting that the derivative at a discontinuous point yields Dirac's delta function, we obtain such a solution as follows:
\begin{align}
  \label{eq:p(phi)_full}
  p(\theta) = 
 \begin{cases}
   \frac{\alpha \exp\left[\alpha
  \left(\frac{1}{\tan\frac{\thdeath}{2}}-\frac{1}{\tan\frac{\thdeath
  - \theta}{2}}\right)\right]}{1-\cos(\thdeath - \theta)} & (0 \leq \theta
  < \thdeath),\\
  0 & (\text{otherwise}),
 \end{cases}
 \end{align}
 which is continuous at $\theta=\thdeath$ but discontinuous at $\theta=0$.

Inserting Eq.~\eqref{eq:p(phi)_full} into Eq.~\eqref{eq:def_v} and imposing $v(\thdeath)=\frac{\dd v}{\dd \theta}(\thdeath)=0$, we find that the parameters $\kappa$ and $\alpha$ satisfy
\begin{align}
 \kappa =
g_1\left(\cth(\alpha);\alpha\right)\eqqcolon \kdeath\left(\alpha\right),\label{eq:kappa_death}
\end{align}
where
\begin{align}
  g_1(c; \alpha) &\coloneqq
   \left(
   2 \alpha e^{\frac{\alpha c}{\sqrt{1 - {c}^2}}} 
   \int_{c}^{1} 
   \frac{x e^{\frac{-\alpha x}{\sqrt{1 - x^2}}}}{1-x^2} \dd x
   \right)^{-1}, \label{eq:g1}
\end{align}
and $\cth \coloneqq \cos \frac{\thdeath}{2}$ is a zero of the function
\begin{align}
 g_2(c;\alpha) =
 \int_c^{1} 
 \frac{(2x^2 - 1)
 e^{\frac{-\alpha x}{\sqrt{1-x^2}}}
 }
 {\left(1-x^2\right)^{3/2}} \dd x. \label{eq:g2}
 \end{align}

We solve $g_2=0$ numerically and insert the solutions into Eq.~\eqref{eq:kappa_death} to obtain $\kdeath$, which is plotted as a yellow solid curve in \figref{fig:Q}(b).
The curve agrees with the value of $\alpha$ at which $Q$ vanishes for large $\kappa$, supporting the assertion that the disappearance of the collective oscillation is due to SOQ.

\paragraph{Conclusion}
  This study has investigated a model of coupled oscillators with turnover. In contrast to a previous study \cite{Zwicker2010} that focused solely on turnover, we have highlighted the synergistic effect between coupling and turnover. As a result, the collective oscillation has been shown to disappear via two types of transitions depending on both the coupling strength and the turnover rate; one is desynchronization, and the other is a novel type of transition termed SOQ. Importantly, SOQ cannot be induced by turnover or coupling alone, but by their synergistic effect.

  Characterizing these two distinct types of transitions has potential implications for the control and design of the collective behavior of oscillatory assemblies. 
First, it helps avoid the unexpected disappearance of synchronous oscillations.  In the original Kuramoto model, increasing the coupling strength enhances phase cohesion and promotes the collective oscillation, so one would expect the same to be true for oscillators with turnover.
However, if the turnover rate is sufficiently large, increasing the coupling strength causes the collective oscillation to disappear through SOQ.
 Second, the type of transition may be exploited for designing appropriate forcing to steer the system to a desirable state, as has been done in Ref.~\cite{Murayama2017}.

Previous experimental studies have proposed various 
open reactors that sustain non-equilibrium chemical reactions by controlling the fluxes into and out of the system \cite{Epstein1989,Spirin2004,Sugiura2016,Niederholtmeyer2013}. 
Desynchronization and SOQ may be observed if self-sustained oscillators are incorporated into such an apparatus and parameters are varied. A candidate system might be constructed by combining techniques to maintain constant protein turnover \cite{Niederholtmeyer2013} with protein molecules whose states change periodically \cite{Kimchi2020, Nakajima2005}.
Our work might also be relevant to tissue formation in multicellular organisms because proliferation and differentiation sometimes involve phase resetting of cellular oscillators \cite{Fukuda2012,Yagita2010}.

  Desynchronization transition has been observed in various coupled oscillator systems. In contrast, whether SOQ occurs robustly in different oscillator systems needs to be investigated in future studies. 
  The investigation of other oscillator models will provide insights into the robustness of the transition \cite{Ozawainpreparation}.

\begin{acknowledgments}
  The authors thank Hiroshi Ito and Keiko Imai for helpful discussions on protein turnover, Ryota Kobayashi for illuminating comments on the Poisson process, and Namiko Mitarai, Tetsuhiro Hatakeyama, and Yuting Lou for insightful discussions on the application of the theory. This study was supported by JSPS KAKENHI Grant Number JP22KJ0899 to A.O. and JSPS KAKENHI Grant Number JP21K12056 to H.K.
\end{acknowledgments}
A.O. and H.K. conceptualized the work; A.O. performed analyses and wrote the manuscript with support from H.K.

\end{document}


\title{Supplemental Material for `Two distinct transitions in a population of coupled oscillators with turnover: desynchronization and stochastic oscillation quenching'}
\author{Ayumi Ozawa}
\email[corresponding author: ]{aozawa@g.ecc.u-tokyo.ac.jp}
\author{Hiroshi Kori}
\affiliation{Graduate School of Frontier Sciences, The University of Tokyo, Chiba 277-8561, Japan}
\date{\today}
\maketitle

\section{Algorithm for simulating Equation (2)}
Equation (2) was simulated using the Euler-Maghsoodi method\cite{Hanson2007} as follows. 
    Let $\Delta t$ be a small time step and $\vartheta_i^{(k)}$ be the value of $\theta_i$ at time $k \Delta t$.
    We update $\vartheta_i^{(k)}$ by
    \begin{align}
      \vartheta_i^{(k+1)} &= 
      \begin{cases}
        \vartheta_i^{(k)} + v_i^{(k)}\Delta t & (\text{if } \Delta P_i^{(k)}= 0),\\
        \varphi_i^{(k)} + v_i^{(k)}\Delta t  & (\text{if } \Delta P_i^{(k)}=1),
      \end{cases} \tag{S1}\\
      v_i^{(k)} &= \omega + \frac{\kappa}{N} \sum_{j=1}^{N} \sin\left( \vartheta_j^{(k)} - \vartheta_i^{(k)}\right), \tag{S2}
    \end{align}
    where $\Delta P_i^{(k)}$ is a pseudorandom number, which corresponds to the change in a sample path of the Poisson process $P_i$ from time $k \Delta t$ to time $(k+1) \Delta t$. 
    In practice, when $\Delta t$ is sufficiently small, $\Delta P_i^{(k)}=1$ with probability $\alpha \Delta t$, and otherwise $\Delta P_i^{(k)}=0$.
    If $\Delta P_i^{(k)}=1$, i.e., when a resetting event occurs, a pseudorandom number $\varphi_i^{(k)}$, which corresponding to the realization of $\phi_i$ at that event, is generated according to the distribution $f(\phi)$ to calculate $\vartheta_i^{(k+1)}$.